\documentstyle[twocolumn,prl,aps]{revtex}
\begin{document}
\title{Motion induced radiation from a vibrating cavity}
\author{Astrid Lambrecht$^a$, Marc-Thierry Jaekel$^b$ and Serge Reynaud$^c$}
\address{$(a)$ Max-Planck-Institut f\"ur Quantenoptik and
Ludwig-Maximilians-Universit\"at M\"unchen,\\
Hans-Kopfermann-Str.1, D-85748 Garching, Germany \\
$(b)$ Laboratoire de Physique Th\'{e}orique de l'Ecole Normale
Sup\'{e}rieure,
Universit\'{e} de Paris-Sud, \\
Centre National de la Recherche Scientifique,
24 rue Lhomond, F-75231 Paris Cedex 05, France\\
$(c)$ Laboratoire Kastler Brossel, Ecole Normale Sup\'{e}rieure,
Universit\'{e} Pierre et Marie Curie, \\
Centre National de la Recherche Scientifique,
4 place Jussieu, F-75252 Paris Cedex
05, France}
\date{LPTENS 96/03  - to appear in Physical Review Letters}
\maketitle

\begin{abstract}
We study the radiation emitted by a cavity moving in vacuum.
We give a quantitative estimate of the
photon production inside the cavity as well as
of the photon flux radiated from the cavity. A resonance
enhancement occurs not only when the cavity length is
modulated but also for a global oscillation of the cavity.
For a high finesse cavity the emitted radiation surpasses
radiation from a single mirror by orders of magnitude.

\end{abstract}

Vacuum field fluctuations exert radiation pressure forces on any scatterer
placed in empty space. For two mirrors at rest in vacuum, this effect
is known since a long time as the Casimir effect \cite{Casimir}. It has more
recently been recognized that dynamical
counterparts of this static force appear for moving scatterers. For some
types of motion, the field does not remain in the vacuum state, but photons
are produced through non adiabatic processes \cite{Moore}. Due to energy
conservation,
the scatterers motion then has to be damped out and this damping may be
associated with dissipative radiation reaction forces.

Motion induced effects of vacuum radiation pressure do not require the
presence of two mirrors but already exist for a single mirror moving in
vacuum. In this case, the radiation reaction force is known to arise as soon
as the mirror has a non-uniform acceleration \cite{FullingDavies}. The
effects of radiation from a moving mirror and the associated radiation
reaction force raise intriguing questions with respect to the standard
mechanical description of motion. They imply that dissipative effects are
associated with the motion of mirrors in vacuum, although this motion has no
further reference than vacuum itself. They thus seem to challenge the
principle of relativity of motion. It would therefore be very important to
obtain experimental evidence of these dissipative processes associated with
motion in vacuum. However vacuum radiation pressure scales as Planck's
constant $\hbar $ and produces therefore only small mechanical perturbations
for any macroscopic mirror, so that the feasibility of an experimental
demonstration of motion induced dissipation is usually considered to
lie out of reach of present technology.

The aim of the present Letter is to show that quantitative figures are greatly
improved when the attention is focussed onto the
emission of radiation from an empty high finesse cavity oscillating in
vacuum. Indeed the number of emitted photons is the ratio of the radiated
energy to the photon energy and thus scales as $\hbar ^0$. This argument
clearly supports a detection of optical rather than mechanical signatures of
motion induced dissipation. Furthermore, a cavity configuration should
allow to take advantage of resonance enhancement effects.

Motion induced
radiation can be interpreted using analogies with optical parametric
processes. It is well known that the cavity field is parametrically excited
when the mechanical cavity length is modulated at a frequency equal to an
even integer multiple of the fundamental optical resonance frequency. If the
cavity field is initially in the vacuum state, this excitation leads to
a squeezed vacuum state \cite{squeezing} which differs from the pure vacuum
state and in particular contains photons. Compared to the situation with a
single oscillating mirror, radiation is resonantly enhanced in this cavity
configuration. More strikingly, a resonant enhancement also exists when the
cavity oscillates as a whole, with its mechanical length kept constant, at
frequencies equal to odd integer multiples of the fundamental optical
resonance frequency. Motional radiation is in this case reminiscent of
photon emission from a single oscillating mirror, however with the
difference that it is enhanced by the cavity finesse.

A number of calculations has been devoted to the energy build-up inside a
cavity with perfect mirrors \cite{LDKM}. However these calculations do not
provide satisfactory answers to the previously discussed questions. They do
not consider the photons radiated by the cavity since the latter is treated
like a closed system. Even for the photons produced inside the cavity, the
hypothesis of perfect mirrors amounts to disregard the important problem of
finite lifetime of photons inside the cavity. In this Letter in contrast, we
study the configuration of a cavity built with partly transmitting mirrors.
The cavity thus appears as an open system able to radiate into the free
field vacuum. At the same time, the influence of the cavity finesse may be
quantitatively evaluated.

For simplicity, we limit ourselves here to two-di\-men\-sio\-nal space-time
calculations. As is well known from the analysis of squeezing experiments,
the transverse structure of the cavity modes does not change appreciably the
results obtained from this simplified model. Each transverse
mode is correctly described by a two-dimensional model as soon as the size
of the mirrors is larger than the spot size associated with the mode.
The two-dimensional model thus corresponds to a conservative estimate where
one transverse mode is efficiently coupled to the moving mirrors. A more
precise evaluation for a realistic configuration should take diffraction
into account and would probably lead to a result obtained by
multiplying the two-dimensional result by the Fresnel number, i.e. the
number of efficiently coupled transverse modes \cite{calculs4D}.

Before studying the cavity
configuration, we consider briefly the case of a single moving mirror and
calculate the photon flux as well as the spectrum of the emitted radiation.
To derive the radiation, we use general arguments associated with scattering
theory, without specific assumptions on the form of the interaction between
mirror and field. This approach does not rely on a detailed microscopic
analysis and is therefore applicable to any type of mirror as long as its
internal dissipation is negligible. We disregard the recoil of the
mirror which is supposed to have a macroscopic mass. To specify the scattering
properties of the
mirror, we introduce column-matrices $\Phi \left[ \omega \right] $ which
contain the components at a given frequency $\omega $ of the free fields
propagating in opposite directions
\begin{eqnarray}
\Phi \left[ \omega \right] =\sqrt{\frac \hbar {2\left| \omega \right| }}%
\left[
\begin{tabular}{c}
$\theta (\omega )a_{+,\omega }+\theta (-\omega )a_{+,-\omega }^{\dagger }$
\\
$\theta (\omega )a_{-,\omega }+\theta (-\omega )a_{-,-\omega }^{\dagger }$%
\end{tabular}
\right]
\end{eqnarray}
{}Field components with positive or negative frequencies correspond
respectively to annihilation ($a_{\pm ,\omega }$) and creation ($a_{\pm
,\omega }^{\dagger }$) operators ($\theta $ is the Heaviside step function).
The transformation from the input field $\Phi _{{\rm {in}}}$ to the output
one $\Phi _{{\rm {out}}}$ is described by a unitary $S$-matrix which
contains the transmission and reflection amplitudes $s\left[ \omega \right] $
and $r\left[ \omega \right] $ at a given frequency
\begin{eqnarray}
\Phi _{{\rm {out}}}\left[ \omega \right] =\left[
\begin{tabular}{cc}
$s\left[ \omega \right] $ & $r\left[ \omega \right] $ \\
$r\left[ \omega \right] $ & $s\left[ \omega \right] $%
\end{tabular}
\right] \Phi _{{\rm {in}}}\left[ \omega \right]
\end{eqnarray}
The scattering of the field on a motionless mirror does not change the
field frequency and the vacuum state is then preserved, as a consequence of
unitarity.

When the mirror is moving, the frequency of the field is changed by the
scattering process and the $S$-matrix now describes this frequency change
\begin{equation}
\Phi _{{\rm {out}}}\left[ \omega \right] =\int \frac{d\omega ^{\prime }}{%
2\pi }S\left[ \omega ,\omega ^{\prime }\right] \Phi _{{\rm {in}}}\left[
\omega ^{\prime }\right]
\end{equation}
Assuming that the incoming field is in the vacuum state, one obtains the
following expression for the number $N$ of photons radiated into vacuum by
the moving mirror
\begin{eqnarray}
N &=&\int_0^\infty \frac{d\omega }{2\pi }\int_0^\infty \frac{d\omega
^{\prime }}{2\pi }n\left[ \omega ,\omega ^{\prime }\right]  \nonumber \\
n\left[ \omega ,\omega ^{\prime }\right] &=&\frac \omega {\omega ^{\prime }}%
{\rm {Tr}}\left( S\left[ \omega ,-\omega ^{\prime }\right] S\left[ \omega
,-\omega ^{\prime }\right] ^{\dagger }\right)
\end{eqnarray}
$n\left[ \omega ,\omega ^{\prime }\right] $ is the spectral density which
describes the number of particles present in the output field. Photon
creation from vacuum is associated with a scattering process from a negative
frequency $-\omega ^{\prime }$ to a positive frequency $\omega$. The
radiation has to be summed over the two output ports as indicates the trace $%
{\rm {Tr}}()$.

When the $S$-matrix is evaluated in a first order expansion in the
displacement, which is valid for small displacements in which we are
interested here, the spectral density $n\left[ \omega ,\omega ^{\prime
}\right]$ is proportional to the square modulus of the frequency component $%
\delta q\left[ \omega +\omega ^{\prime }\right] $ of the displacement
\begin{eqnarray}
n\left[ \omega ,\omega ^{\prime }\right] &=&\frac{\omega \omega ^{\prime }}{%
c^2}\gamma \left[ \omega ,\omega ^{\prime }\right] \left| \delta q\left[
\omega +\omega ^{\prime }\right] \right| ^2  \nonumber \\
\gamma \left[ \omega ,\omega ^{\prime }\right] &=&2(1-s\left[ \omega \right]
s\left[ \omega ^{\prime }\right] +r\left[ \omega \right] r\left[ \omega
^{\prime }\right]  \nonumber \\
&&+1-s\left[ \omega \right] ^{*}s\left[ \omega ^{\prime }\right]
^{*}+r\left[ \omega \right] ^{*}r\left[ \omega ^{\prime }\right] ^{*})
\end{eqnarray}
This expression results from a linear approximation of the motional
perturbation of the field, but it is valid without any restriction on the
motion's frequency. It is directly connected to the general relation which
exists between the motional perturbation of the scattering matrix and the
radiation pressure force exerted upon the mirror \cite{MirrorVacuum}.

In the following, we consider the case of a mirror following a harmonic
motion at a frequency $\Omega $. Since we expect the radiation of photons to
be proportional to time, we focus our attention onto a harmonic motion of
amplitude $a$ during a time $T$
\begin{equation}
\delta q\left( t\right) =2a\cos \left( \Omega t\right) \qquad 0<t<T
\end{equation}
{}For a long oscillation time $T$, we find the number $N$ of radiated photons
to be defined per unit time
\begin{equation}
\frac NT=\frac{a^2}{c^2}\int_0^\Omega \frac{d\omega }{2\pi }\omega \left(
\Omega -\omega \right) \gamma \left[ \omega ,\Omega -\omega \right]
\label{N}
\end{equation}
This result is similar to the expression one would obtain for the number of
photons spontaneously emitted by an atom coupled to vacuum fluctuations,
calculated with Fermi's golden rule. Here the emission is generated by the
parametric coupling of the mirror's mechanical motion to vacuum radiation
pressure rather than by the coupling of the atomic dipole to the vacuum
field. Hence photons are emitted through a two-photon parametric process
rather than through a one-photon process. As is well known, spontaneous
emission is not accompanied by absorption processes because vacuum is the
field ground state. Here the same property entails that photons are only
emitted at frequencies $\omega $ and $\omega ^{\prime }$ smaller than the
frequency $\Omega $ of the mechanical motion. Each parametric process
corresponds to the emission of two photons carrying away an energy $\hbar
\Omega =\hbar \left( \omega +\omega ^{\prime }\right)$, so that the radiated
energy may be obtained as $\frac 12N\hbar \Omega $. This energy corresponds
exactly to the work supplied by the mirror against the radiation reaction
force, in absence of other dissipative mechanisms. This consistency between
mechanical dissipation and optical radiation is ensured by expression (\ref
{N}) where $N$ appears to be proportional to the noise spectrum at frequency
$\Omega $ of the fluctuations of vacuum radiation pressure experienced by
the mirror \cite{MirrorVacuum}.

In the limiting case of a nearly perfect mirror ($s\rightarrow0;r\rightarrow
-1$), we obtain a simplified expression for the number of radiated photons
\begin{eqnarray}
\frac NT &=&\frac{8a^2}{c^2}\int_0^\Omega \frac{d\omega }{2\pi }\omega
\left( \Omega -\omega \right) =\frac{2a^2\Omega ^3}{3\pi c^2}  \nonumber \\
N &=&\frac{\Omega T}{6\pi }\left( \frac vc\right) ^2\qquad v=2\Omega a
\label{N1}
\end{eqnarray}
Expression (\ref{N1}) for $N$ is a product of two dimensionless factors,
namely the number of mechanical oscillation periods during the time $T$ and
the square of the maximal velocity $v$ of the mirror divided by the velocity
of light $c$. A characteristic feature of motion induced radiation, which
could be used in an experiment to distinguish it from spurious effects, is
the parabolic shape of its spectral density with a maximum at $\omega =\frac
\Omega 2$.

The derivation of motion induced radiation is similar in the case of two
moving mirrors. Assuming the two mirrors to follow a harmonic motion at the
same frequency $\Omega $ with respective amplitudes $a_i$ ($i=1,2$), we
deduce the number of photons radiated per unit time to be
\begin{equation}
\frac NT=\sum_{ij}\frac{a_ia_j}{c^2}\int_0^\Omega \frac{d\omega }{2\pi }%
\omega \left( \Omega -\omega \right) \gamma _{ij}\left[ \omega ,\Omega
-\omega \right]  \label{rad2}
\end{equation}
As in the case of a single mirror, the functions $\gamma _{ij}$ already
appear in the evaluation of motional forces and they have been studied
previously \cite{Motional}. We introduce here simplifying assumptions
allowing to obtain analytical expressions for the motional radiation. In the
frequency range $[0,\Omega ]$ one can in a good approximation assume the
reflexion coefficients $r_1$ and $r_2$ of the two mirrors to be real and
frequency independent. In the following we are concentrating on the most
interesting case where the cavity has a high finesse which implies that both
$r_1$ and $r_2$ are close to unity. Since the functions $\gamma _{ij}\left[
\omega ,\omega ^{\prime }\right] $ exhibit resonances when one of the
emission frequencies $\omega $ or $\omega ^{\prime }$ corresponds to a
cavity mode, we will keep the reflection coefficients, which appear in their
denominators and thus determine their resonant behavior. In contrast, we
will set to unity the reflection coefficients acting only as weighting
factors in the numerators. With these assumptions the functions $\gamma _{ij}
$ only depend on the product $r_1r_2$ of the reflection coefficients which
we denote
\begin{equation}
r_1r_2=e^{-2\rho }\qquad \rho \ll 1
\end{equation}
where $\frac 1\rho $ measures the cavity finesse. They then read
\begin{eqnarray}
&&\gamma _{11}\left[ \omega ,\omega ^{\prime }\right] =\gamma _{22}\left[
\omega ,\omega ^{\prime }\right] =4+4D_{+}\left[ \omega \right] D_{+}\left[
\omega ^{\prime }\right]  \label{defgamma} \\
&&\gamma _{12}\left[ \omega ,\omega ^{\prime }\right] =\gamma _{21}\left[
\omega ,\omega ^{\prime }\right] =-4D_{-}\left[ \omega \right] D_{-}\left[
\omega ^{\prime }\right]  \nonumber \\
&&D_{+}\left[ \omega \right] =\frac{\sinh (2\rho )}{\cosh (2\rho )-\cos
(2\omega \tau )}=\sum_{k=-\infty }^\infty \frac \rho {\rho ^2+\left( \omega
\tau -k\pi \right) ^2}  \nonumber \\
&&D_{-}\left[ \omega \right] =\frac{2\sinh (\rho )\cos (\omega \tau )}{\cosh
(2\rho )-\cos (2\omega \tau )}=\sum_{k=-\infty }^\infty \frac{(-1)^k\rho }{%
\rho ^2+\left( \omega \tau -k\pi \right) ^2}  \nonumber
\end{eqnarray}
$\tau $ is the time of flight of a photon from one mirror to the other. With
exception of the first term in $\gamma _{11}$, all terms contain
denominators clearly associated with the presence of the cavity.

We can now calculate the emitted photon number by performing the integration
(\ref{rad2}) for the various Lorentzian components of the spectrum. Using
the assumption of a high finesse cavity we find
\begin{eqnarray}
\frac NT &=&\frac{\Omega ^3(a_1^2+a_2^2)}{3\pi c^2}+\sum_{k,k^{\prime
}=1}^\infty \frac{N_{k,k^{\prime }}}T  \nonumber \\
\frac{N_{k,k^{\prime }}}T &=&\frac 1\tau \frac{k\pi }{c\tau }\frac{k^{\prime
}\pi }{c\tau }\frac{4\rho \left( a_1-(-1)^{k+k^{\prime }}a_2\right) ^2}{%
4\rho ^2+\left( \Omega \tau -k\pi -k^{\prime }\pi \right) ^2}
\label{Ngeneral}
\end{eqnarray}
The photon flux outside the cavity can also be written by resumming the
contributions of all modes
\begin{eqnarray}
\frac NT &=&\frac{\Omega ^3}{3\pi c^2}(a_1^2+a_2^2)  \nonumber \\
&&+\frac \Omega {6\pi c^2}(\Omega ^2-\frac{\pi ^2}{\tau ^2})\frac{\sinh
(2\rho )(a_1+a_2)^2}{\cosh (2\rho )+\cos (\Omega \tau )}  \nonumber \\
&&+\frac \Omega {6\pi c^2}(\Omega ^2-\frac{\pi ^2}{\tau ^2})\frac{\sinh
(2\rho )(a_1-a_2)^2}{\cosh (2\rho )-\cos (\Omega \tau )}  \label{Ngeneral2}
\end{eqnarray}
The first term in these expressions is a non-resonant contribution coming
from direct reflection of vacuum fluctuations on both sides of the cavity.
All other terms describe resonances of the motional radiation occuring when
the mechanical excitation frequency $\Omega $ is close to an integer
multiple of the fundamental optical resonance frequency $\frac \pi \tau $. $%
N_{k,k^{\prime }}$ describes parametric emission of radiation into the
optical modes of frequencies $k\frac \pi \tau $ and $k^{\prime }\frac \pi
\tau $. Compared to the result obtained for a single mirror, the radiated
photon flux is enhanced by a resonance factor which is essentially the
cavity finesse. For the lowest mechanical resonance at $\Omega =\frac{2\pi }%
\tau $, only one intracavity mode is excited ($k=k^{\prime }=1$). This
corresponds to the situation studied in most works on intracavity field
build-up \cite{LDKM}. In the more general frame developed in the present
Letter, higher resonance frequencies exist, giving rise to several emission
peaks. The emission peaks all have the same spectral width given by the
cavity finesse and their relative intensities reproduce a parabolic
spectrum, as the one obtained for a single moving mirror, however with a
large resonant enhancement. The information contained in the set of peaks
can again be used to distinguish motion induced radiation from spurious
effects.

In equation (\ref{Ngeneral}), even modes $\Omega =\frac{2\pi }\tau ,\frac{%
4\pi }\tau \ldots $ appear as elongation modes which correspond to a
periodic modulation of the mechanical cavity length. In contrast, odd modes $%
\Omega =\frac{3\pi }\tau ,\frac{5\pi }\tau \ldots $ are excited by a global
translation of the cavity with its length kept constant. The latter
effect is thus reminiscent of radiation of a single oscillating mirror,
since the cavity moves in vacuum without any further reference than vacuum
itself. However radiation is now enhanced by the cavity finesse. These
two kinds of vibration modes, which appear to be contrasted in a mechanical
point of view, have been obtained in a unified manner in our scattering
approach which deals with the field bouncing back and forth in the cavity.
The basic reason for this similar description within the scattering formalism
is that the optical length as seen by the field varies in the same way for
both kinds of modes, although the mechanical cavity length is modulated in
one case and constant in the other one.

To estimate the stationary number of photons inside the cavity, we may use a
simple balance argument. Each photon has a probability $4\rho $ of escaping
from the cavity during each roundtrip time $2\tau $. As we know the photon
flux emitted by the cavity per unit time, we can deduce the number of
photons ${\cal N}_{k,k^{\prime }}$ produced by the oscillation in a pair of
cavity modes
\begin{equation}
{\cal N}_{k,k^{\prime }}=\frac{k\pi }{c\tau }\frac{k^{\prime }\pi }{c\tau }%
\frac{2\left( a_1-(-1)^{k+k^{\prime }}a_2\right) ^2}{4\rho ^2+\left( \Omega
\tau -k\pi -k^{\prime }\pi \right) ^2}  \label{Ninside}
\end{equation}

So far we have given a quantitative estimate for the number $N$ of radiated
photons as well as for the number ${\cal N}$ of photons produced inside the
cavity. The model of a cavity with partly transmitting mirrors allows us to
evaluate now the resonance enhancement factor in terms of the cavity
finesse. A mechanical excitation at exact resonance leads to the following
orders of magnitude for $N$ and ${\cal N}$
\begin{eqnarray}
N &\simeq &\frac{\Omega T}{2\pi }\frac{v^2}{c^2}\frac 1\rho   \nonumber \\
{\cal N} &\simeq &\frac{v^2}{c^2}\frac 1{\rho ^2}  \label{Nfinesse}
\end{eqnarray}
where $v$ measures either the sum or the difference of the peak velocity of the
vibrating mirrors, depending on the mode
parity. We may emphasize that not only the number of photons inside the
cavity, but also the number of radiated photons, diverge at the limit of
perfectly reflecting mirrors where the finesse of the optical and mechanical
resonances goes to infinity. This shows that the simple model which treats
the cavity as a closed system misses important physical phenomena.

To be more specific about the orders of magnitude, let us recall that we
have assumed the input fields to be in the vacuum state. This assumption
requires the number of thermal photons per mode to be smaller than $1$ in
the frequency range of interest ($\hbar \omega <k_B\Theta $ with $k_B$ the
Boltzmann constant and $\Theta $ the temperature). Low temperature
technology thus points to experiments using small mechanical structures
with optical resonance frequencies as well as mechanical oscillation
frequencies in the {\rm GHz} range. In this frequency range, the finesse of
a superconducting cavity can reach $10^9$ \cite{Rydberg}. A peak velocity $%
v\simeq 1$~{\rm m/s}, corresponding to an amplitude in the {\rm nm} range,
would thus be sufficient to obtain a radiated flux of $10$ photons per
second outside and a stationary number of $10$ photons inside the cavity. It
is important to emphasize that the peak velocity considered in the present
analysis is only a small fraction of the typical sound velocity in materials
so that fundamental breaking limits do not oppose to these numbers. The
photons may be detected outside the cavity by performing sensitive
photon-counting detection of the radiated flux. Inside the cavity the state
of the field could be probed with the help of Rydberg atoms \cite{Rydberg}.
Therefore, if a technique is found to excite a vibrating motion with the
above characteristics, the challenge of an experimental observation of
motional radiation in vacuum can be taken up.

{\bf Acknowledgements} We would like to thank M.~Devoret, D.~Est\`{e}ve,
T.W.~H\"{a}nsch, S.~Haroche, P.A.~Maia Neto and J.-M.~Raimond for useful
discussions.

\end{document}